\documentclass{pramana}

\usepackage{graphicx,amsmath,bm}
\usepackage{titlesec}

\titleformat{\section}
{\normalfont\fontsize{12}{17}\sffamily\bfseries}
{\thesection}
{1em}
{}

\titleformat{\subsection}
{\normalfont\fontsize{12}{17}\sffamily\bfseries\slshape}
{\thesubsection}
{1em}
{}

\usepackage{placeins}
\usepackage[  
colorlinks=true,
pdfborder={0 0 0},
citecolor=blue,
linkcolor=blue
]{hyperref}
\newcommand*{\myeqref}[2][Equation~]{%
	\hyperref[{#2}]{#1(\ref*{#2})}%
}
\def\equationautorefname#1#2\null{%
	Eq.#1(#2\null)%
}

\begin{document}

%%paper title
%%For line breaks \\ can be used within title 
\title{Conservation Laws for the Nonlinear Klein-Gordon Equation in $\left(1+1\right)-, \left(2+1\right)-,$ and $\left(3+1\right)-$dimensions.}

%%author names are separated by comma (,) 
%%use \and before the last author name 
%%\textsuperscript{number} is used for affiliation
%%use a * along with the number separated by comma
%% for the  author for correspondence

\author{M.A.Z. Khan \textsuperscript{1}, M.Moodley \textsuperscript{2} \and F.Petruccione \textsuperscript{3}}
\affilOne{\textsuperscript{1,2,3} Quantum Research Group, University of KwaZulu-Natal, Private Bag X54001, Durban, South Africa}
%%escape two column mode for title, affiliation and abstract
%%by giving \twocolumn command as shown

\twocolumn[{

\maketitle

%%include \corres to print the corresponding author Email id
\corres{muhammadalzafark@gmail.com}

%%include \msinfo for
%%manuscript information such as
%%received, revised and accepted dates
%%
%\msinfo{}

%%abstract
\begin{abstract}
	We study soliton solutions to the Klein-Gordon equation via Lie symmetries and the travelling-wave \textit{ansatz}. It is shown, by taking a linear combination of the spatial and temporal Lie point symmetries, that soliton solutions naturally exist, and the resulting field lies in the complex plane. We normalize the field over a finite spatial interval, and thereafter, specify one of the integration constants in terms of the other. Solutions to a specific type of nonlinear Klein-Gordon equation are studied via the sine-cosine method, and a real soliton wave is obtained. Lastly, the multiplier method is used to construct conservation laws for this particular nonlinear Klein-Gordon equation in $\left(3+1\right)$-dimensions. 
\end{abstract}
%%insert keywords separated by comma using \keywords{words}

%%include \pacs{number} to print the PACS number
%\pacs{12.60.Jv; 12.10.Dm; 98.80.Cq; 11.30.Hv}

}]
%%close the twocolumn escape here

%%include \doinum{number}for the DOI number in the header
%%include \volnum{number} for the volume number in the header
%%include \year{yyyy} for  year of publication in the header
%%include \pgrange{num--num} page range of article in the header
%%include \artcitid{num} for the article citation id
%%include \lp to print last page of the article
%%include \setcounter{page}{pagenum} for the exact starting page of the article

\doinum{12.3456/s78910-011-012-3}
\artcitid{\#\#\#\#}
\volnum{123}
\year{2016}
\pgrange{1-9}
\setcounter{page}{1}
\lp{9}

\section{Introduction}

The Klein-Gordon (KG) equation is a quantized version of the relativistic dispersion relation for a free particle
\begin{equation}
	E^{2} = (pc)^{2} + (m_{0}c^{2})c^{2}, \tag{1}\label{1}
\end{equation}
where $E$ denotes the particle energy; $p$, the relativistic momentum; $c$, the speed of light; and $m_{0}$, the rest mass. The KG equation was originally developed to investigate relativistic electrons, but it subsequently became more applicable to spinless relativistic particles such as pions -- ephemeral mesons that can take the place of an electron in an atom creating pionic atoms \cite{ref1}, up and anti-down $\pi^{+}$, down and anti-up $\pi^{-}$, the neutral quark $\pi^{0}$, and the Higgs particle, $H^{0}$ \cite{QFT1,QFT2}. It is considered to be the first relativistic wave equation. \\

The KG equation is Lorentz-covariant; for some arbitrary transformation in the spatial coordinates, $\mathbf{r} \rightarrow \mathbf{r^{\prime}}$, and the temporal coordinate, $t \rightarrow t^{\prime}$, that satisfies the Lorentz group of transformations \cite{ref3}, we have that $\Psi(\mathbf{r},t) = \Psi(\mathbf{r}^{\prime},t^{\prime})$, where $\Psi(\mathbf{r},t)$ is the scalar field that satisfy the KG equation
\begin{equation}
	-\frac{1}{c^{2}}\Psi_{tt}(\mathbf{r},t)=\left(-\bm{\nabla}\cdot\bm{\nabla}+\frac{m^{2}c^{2}}{\hbar^{2}}\right)\Psi(\mathbf{r},t). \tag{2}\label{2}
\end{equation}
In \myeqref{2}, $\hbar$ denotes the reduced Planck constant, and the subscripts denote partial differentiation. We provide a derivation of \myeqref{2}, employing the principle of least action, in \autoref{EOM} below. \\ 

The KG equation was derived independently by Klein \cite{Klein}, Gordon \cite{Gordon}, and Schr\"{o}dinger \cite{Schroedinger} as a relativistic extension of Schr\"{o}dinger's equation. Initially, the solutions were disregarded, because they lead to a non-definite positive probability. The probability current
\begin{equation}
	J_{\mu}=\Psi^{*}(\mathbf{r},t)\partial_{\mu}\Psi(\mathbf{r},t)-\left[\partial_{\mu}\Psi^{*}(\mathbf{r},t)\right]\Psi(\mathbf{r},t), \tag{3}\label{3}
\end{equation}
leads to the four-divergence being zero, and thus, the time-component of the probability density current cannot be interpreted. General solutions to the KG equation are composed of a superposition of two states, namely, the incoming positive energy particle and the outgoing positive energy antiparticle. The general solution is therefore not an inherently single-particle description, because one is forced to consider particles and antiparticles \cite{ref14}. \\

The Lorentz-covariance naturally implies rotatio-nal-invariance, and therefore, $\Psi(\mathbf{r},t)$ represents spin-zero particles. Another feature of the Lorentz-covari-ance is that $\vert \Psi \vert ^{2}$ is strictly positive; thus, the energy density, $\rho = 2E\vert \Psi \vert^{2}$, need not be positive.  According to Feynman, the negative four-momentum solutions are used to describe antiparticles; see \cite{ref1, ref3, ref14,ref2, ref5, ref16}. \\

Several researchers have studied solutions to the KG equation, and more general relativistic wave equations, in terms of Lie point symmetries for different scenarios. Paliathanasis \textit{et al} \cite{Paliathanasis} successfully provided a classification scheme for Bianchi I spacetimes and thereafter, provided generalized results that related the Lie point symmetries to the conformal algebras of the underlying geometry. Jamal \textit{et al} \cite{JamalKaraBokhari} used Lie point symmetries to classify the KG equation and thereafter, calculated exact solutions and conservation laws in de Sitter spacetimes. Azad \textit{et al} \cite{Azad} performed symmetry analysis, Iwasawa algebraic decomposition, and algebra classification using maximal isometry groups corresponding to Minkowski, de Sitter, anti-de Sitter, Einstein, anti-Einstein, Bertotti-Robinson, and other metrics of the Petrov type D for static spherically symmetric spacetimes. Paliathanasis and Tsamparlis \cite{PaliathanasisTsamparlis} calculated the Lie point symmetries for the KG equation in a general Riemannian space and showed its relation to homothetic and conformal algebras. In \cite{Inc}, Inc \textit{et al} studied the time-fractional KG equation (f-PDE) and demonstrated its reduction to a fractional nonlinear ODE (fn-ODE) upon which a closed-form solution was provided in terms of a power series. \\

Furthermore, several authors have studied soliton solutions to the KG equation. In \cite{Biswas}, Biswas \textit{et al} studied the perturbed and unperturbed KG equation under log-law and power-law nonlinearities, producing soliton solutions and conservation laws using the Lie transformation groups. Khalique and Biswas \cite{KhalqiueBiswas} use the travelling-wave \textit{ansatz} to obtain explicit soliton solutions for the nonlinear KG equation, with the nonlinearity assuming the form of the the gradient of some arbitrary potential function. Wazwaz \cite{Wazwaz1}, the originator of the hyperbolic tangent (tanh) and sine-cosine methods described in \autoref{TLKGE} and \autoref{TNLKGE} below, successfully applied these techniques to the KG equation and the nonlinear KG equation under power-law nonlinearity, producing a whole class of soliton solutions. \\

Mathematically, the KG equation has the property that it is a linear homogeneous partial differential equation (PDE) with constant coefficients. This feature is particularly paramount, because it implies infinitely many conservation laws. Therefore, to determine the conservation laws would prove to be a superfluous task. However, if one had to modify the KG equation so that it becomes nonlinear, then, conservation laws become important since these quantities are not well known. \\ 

Besides quantum field theory, several researchers have applied the KG equation to examine phenomena in classical mechanics, astrophysics and cosmology, particle physics, nonlinear optics, solid state physics (in particular, dislocations in crystals), and wave phenomena \cite{ref1, ref11, ref12, Kragh}. \\

In this paper, we demonstrate, from the nature of the Lie point symmetries, soliton solutions naturally arise. For the linear KG equation, we easily obtain soliton solutions using the travelling-wave $\textit{ansatz}$, recovering the known result obtained by Waz-waz \cite{Wazwaz1}. Soliton solutions to the nonlinear KG equation are then studied using the sine-cosine meth-od and an explicit solution is presented. Lastly, we calculate conservation laws using the multiplier method. This paper is organized as follows: In \autoref{EOM}, we derive the underlying equation of motion using the calculus of variations, and specifically, the principle of least action. In \autoref{TLKGE}, we demonstrate the solutions that arise from the Lie point symmetries of the KG equation, and thereafter, obtain the soliton solution to the linear KG equation. We calculate the exact solution for the nonlinear KG equation in \autoref{TNLKGE}, and thereafter, calculate the associated conservation laws using the multiplier method. A conclusion to this paper is presented in \autoref{CONCLUSION} wherein the results obtained are reflected upon. The appendices provide a brief discussion of Lie symmetry analysis, the multiplier method, solitons, and the sine-cosine method.

\section{Equations of Motion}\label{EOM}
Here, we derive the KG equation using the principle of least action. Consider the scalar field,  $\Psi(\mathbf{r},t)\in\mathbb{C}$, whose Lagrangian density is given by
\begin{equation}
	\mathcal{L}=\frac{1}{c^{2}}\Psi_{t}^{2} - (\bm{\nabla}\Psi)^{2} - \dfrac{m^{2}c^{2}}{\hbar^{2}}\Psi^{2}. \tag{4}\label{4}
\end{equation}
This Lagrangian is the relativistic analogue of the harmonic oscillator, encountered in classical mechanics, and describes the motion of massive scalar fields. Furthermore, an external source, $\mathcal{J}(\mathbf{r})$, may be added to \myeqref{4} to describe field interactions. See \cite{ref1} for a full elucidation. Consider the map, $\Psi : \mathbb{R}^{4} \rightarrow \mathbb{R}$. We examine the action
\begin{equation} 
	S[\Psi] = \int_{\mathcal{R}\subset \mathbb{R}^{4}}^{} \mathrm{d}^{4}x \frac{1}{2} \Bigg[\frac{1}{c^{2}}\Psi_{t}^{2} - (\bm{\nabla}\Psi)^{2} - \dfrac{m^{2}c^{2}}{\hbar^{2}}\Psi^{2}\Bigg], \tag{5}\label{5}
\end{equation}
where $\mathcal{R} = \{(t,x,y,z) \vert t_{1}\leq t \leq t_{2}, -\infty <x,y,z<+\infty\}$ is a subspace of the four-dimensional Euclidean space. Varying \myeqref{5} and applying integration by parts, we obtain the KG equation 
\begin{equation}
	-\frac{1}{c^{2}} \Psi_{tt}(\mathbf{r},t) + \nabla^{2}\Psi(\mathbf{r},t) - \dfrac{m^{2}c^{2}}{\hbar^{2}} \Psi(\mathbf{r},t) = 0. \tag{6.1}\label{6.1} 
\end{equation}
In (1+1)-dimensions, \myeqref{6.1} reduces to
\begin{equation}
	\frac{1}{c^{2}}  \Psi_{tt}(\mathbf{r},t) - \Psi_{xx}(\mathbf{r},t) + \dfrac{m^{2}c^{2}}{\hbar^{2}}  \Psi(\mathbf{r},t) = 0. \tag{6.2}\label{6.2}
\end{equation}

\myeqref{6.2} is the original linear KG equation derived independently by Klein, Gordon, and Schr\"{o}dinger. Additionally, it is indicated in \cite{ref1} that V. Fock, J. Kudar, T. de Donder, and H. van Dungen also arrived at \myeqref{6.2}, while attempting to make Schr\"{o}dinger's equation consistent with Einstein's special relativity. In an essay on the KG equation, Kragh \cite{Kragh} quotes a 1926 letter by Pauli to Schr\"{o}dinger \cite{Pauli} in which he describes the KG \myeqref{6.2} as being an \textit{``equation with many fathers''}, alluding to the numerous $20^{\text{th}}$-century theoretical physicists who independently derived the equation. For a justification of the action detailed in \myeqref{5}, see \cite{ref14}.

%==========================================================================================

\section{The Linear Klein-Gordon Equation}\label{TLKGE}
Using $\texttt{MATHEMATICA}$ \cite{ref15}, and applying the methods described in Appendix A, we find that the Lie point symmetries of the linear KG \myeqref{6.2} are $(4+\infty)$-dimensional, that is, four of the symmetries are finite and continuous, and one symmetry in infinite and continuous. Upon solving the system of linear PDEs that arise from the application of \myeqref{A.3.1} on \myeqref{6.2}, we observe that the resulting set of Lie point symmetries imply spatial, temporal, and Galilean boost invariance. \autoref{table:1} below shows the symmetries and corresponding invariant solutions.

\FloatBarrier 
\begin{center}
	\scriptsize
\begin{table}[H]
	\centering
\caption{Some Lie point symmetries of the KG equation where $\gamma_{i}\in\mathbb{R}$ are constants for $1\leq i\leq 5$}
\label{table:1}
\renewcommand{\arraystretch}{1.5}
\hspace{-1cm}	\begin{tabular}{ll}
		\hline
		\textbf{Symmetry} &\textbf{Invariant Solutions} \\ 
		\hline
		$\Omega_{1} =\partial_{x}$ & $\Psi(x,t) = \gamma_{1}\cos\Big(\frac{mc^{2}}{\hbar}t\Big) + \gamma_{2}\sin\Big(\frac{mc^{2}}{\hbar}t\Big)$ \\ 
		\hline
		$\Omega_{2} = \partial_{t}$  & $\Psi(x,t) = \gamma_{3}e^{mcx/\hbar}+\gamma_{4}e^{-mxc/\hbar}$ \\ 
		\hline
		$\Omega_{3} = \frac{1}{c^{2}}x\partial_{t} + t\partial_{x}$  &  $\Psi(x,t) = 2\gamma_{5}\sum\limits_{m=0}^{\infty} \frac{1}{m!\Gamma(m+1)}\Bigg(\frac{mc\sqrt{x^{2}-c^{2}t^{2}}}{2\hbar}\Bigg)^{2m}$ \\
		&$ + \int_{0}^{\infty}\dfrac{1}{\sqrt{\nu^{2} + 1}} \cos \Bigg(\frac{mc\sqrt{x^{2}-c^{2}t^{2}}}{\hbar}\nu\Bigg) \mathrm{d}\nu$ \\
		\hline
	\end{tabular} 
\end{table}
\end{center}
\FloatBarrier

\begin{figure}[H]
	\centering
	\includegraphics[width=0.8\linewidth]{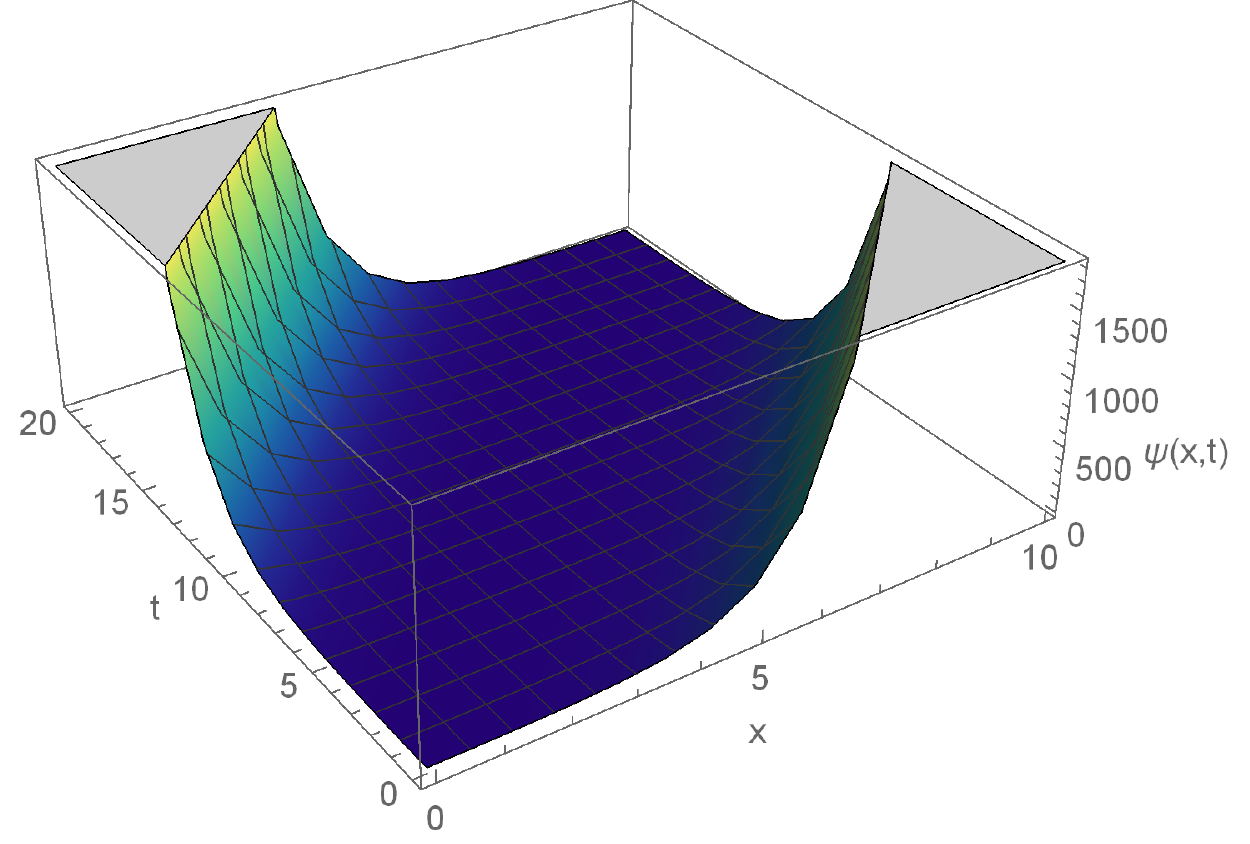}
	\caption{Plot of the travelling wave solution with $A_{1}, A_{2},$ and $m=1$ on the interval $\left(1,3\right)$ in relativistic units that obey causality with wave speed, $w=\frac{1}{2}$}
	\label{fig:10}
\end{figure}

\begin{figure}[H]
	\centering
	\includegraphics[width=0.5\linewidth]{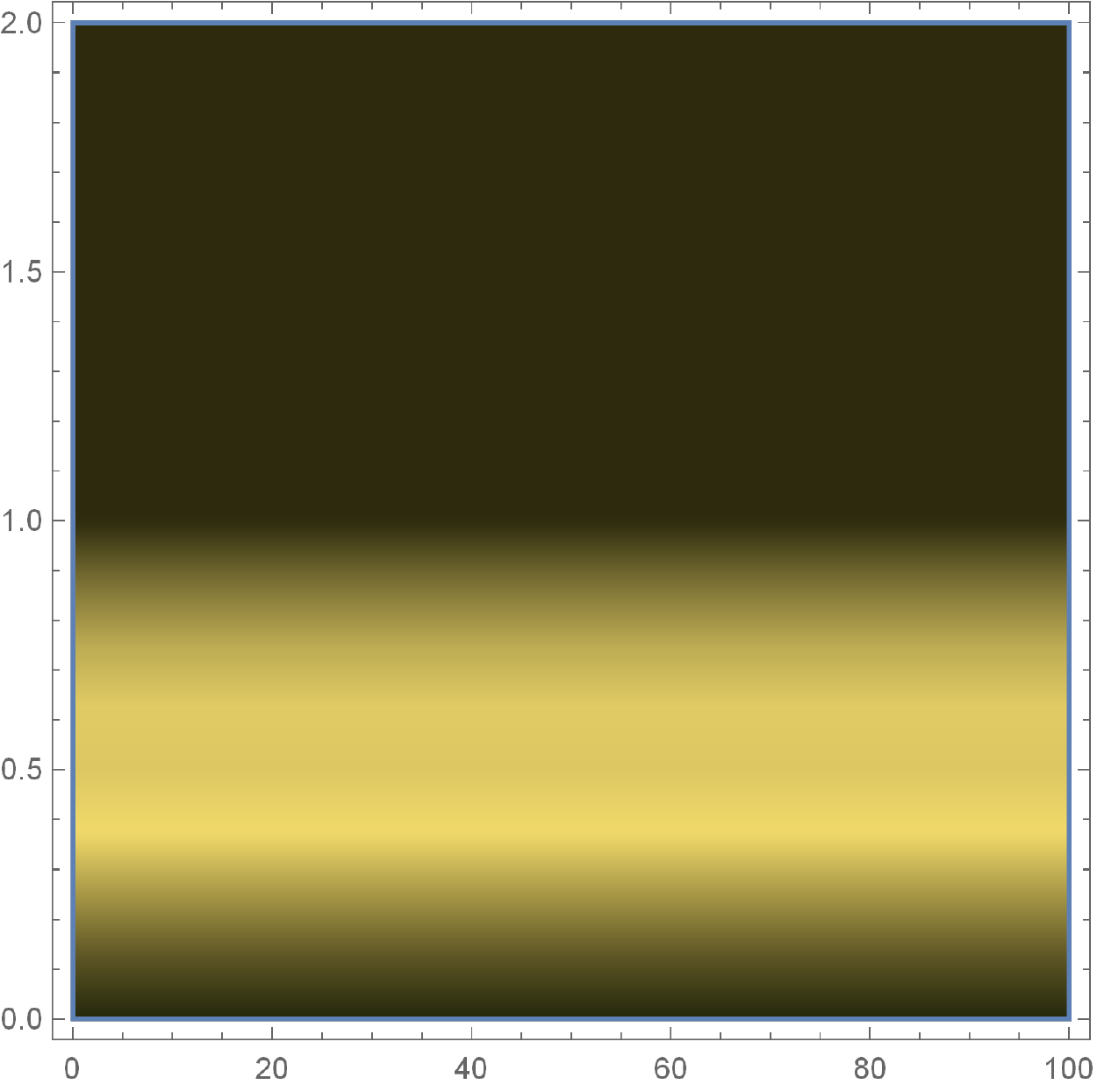}
	\caption{Density plot of $|\Psi(x,t)|$ with $A_{1}, A_{2}$, and $m=1$ on the interval $\left(1,3\right)$ in relativistic units that violate causality with wave speed, $w=2$}
	\label{fig:11}
\end{figure}

%\begin{table}[H]
%	\centering
%	\caption{Some Lie point symmteries of the KG equation where $\gamma_{i}\in\mathbb{R}$ are constants for $1\leq i\leq 5$}
%	\begin{tabularx}{\textwidth}{|l|X|}
%		\hline
%		\hline
%		\hspace*{0.5cm}\textbf{Symmetry} & \hspace*{5cm}\textbf{Invariant Solution} \\
%		\hline
%		$\Omega_{1} =\partial_{x}$ & $\Psi(x,t) = \gamma_{1}\cos\Big(\frac{mc^{2}}{\hbar}t\Big) + \gamma_{2}\sin\Big(\frac{mc^{2}}{\hbar}t\Big)$ \\ 
%		\hline
%		$\Omega_{2} = \partial_{t}$  & $\Psi(x,t) = \gamma_{3}e^{mcx/\hbar}+\gamma_{4}e^{-mxc/\hbar}$ \\ 
%		\hline
%		$\Omega_{3} = \frac{1}{c^{2}}x\partial_{t} + t\partial_{x}$  &  $\Psi(x,t) = 2\gamma_{5}\sum\limits_{m=0}^{\infty} \ddfrac{1}{m!\Gamma(m+1)}\Bigg(\ddfrac{mc\sqrt{x^{2}-c^{2}t^{2}}}{2\hbar}\Bigg)^{2m}$ \newline
%		$\hspace*{1.6cm}+ \int_{0}^{\infty}\dfrac{1}{\sqrt{\nu^{2} + 1}} \cos \Bigg(\ddfrac{mc\sqrt{x^{2}-c^{2}t^{2}}}{\hbar}\nu\Bigg) \mathrm{d}\nu$ \\
%		\hline
%	\end{tabularx}
%\end{table}
The solutions in \autoref{table:1} do not provide any meaningful interpretation. This is because they are either only time-dependent or position-dependent, or written as an infinite series in terms of special functions, namely, the gamma function. As a result, we turn to a soliton approach \cite{ref22,ref23}. Taking a linear combination, $\Omega_{1}-w\Omega_{2}$, of the symmetries, where $\textit{w}$ is the wave speed, we define the invariants
\begin{align} 
	\Xi & := x-wt,  \tag{7.1}\label{7.1} \\
	\phi & := \Psi. \tag{7.2}\label{7.2}
\end{align}
Substituting \myeqref{7.1} and \myeqref{7.2} into \myeqref{6.2}, we obtain the reduced second-order linear ODE
\begin{equation}
	\Bigg[\Big(\frac{w}{c}\Big)^{2}-1\Bigg] \dfrac{\mathrm{d}^{2}\phi}{\mathrm{d}\Xi^{2}} + \frac{m^{2}c^{2}}{\hbar^{2}}\phi = 0. \tag{8}\label{8}
\end{equation}
Solving \myeqref{8} and reintroducing the original variables, we obtain an expression for the scalar field
\begin{equation}
	\Psi(x,t) = A_{1}\exp\Bigg[\frac{mc^{2}(x-wt)}{\hbar\sqrt{c^{2}-w^{2}}}\Bigg] + A_{2}\exp\Bigg[-\frac{mc^{2}(x-wt)}{\hbar\sqrt{c^{2}-w^{2}}}\Bigg], \tag{9}\label{9}
\end{equation}
where $A_{1}$ and $A_{2}$ are the constants of integration. \myeqref{9} is the same solution obtained by Wazwaz \cite{Wazwaz1} (for his $n=1$ case), who studied a generalization of the KG \myeqref{6.2} using the sine-cosine method. Normalizing \myeqref{9} over the interval $\left(a,b\right)\subset\mathbb{R}$, we obtain

\begin{align}
	A_{1} = \frac{mc^{2}}{\lambda\hbar\sqrt{c^{2}-w^{2}}}\left\{2A_{2}\left(b-a\right)\right\} & - \nonumber \\ \left\{\sqrt{\left(2aA_{2}-2bA_{2}\right)^{2} - \frac{\lambda\bar{\lambda}\hbar\sqrt{c^{2}-w^{2}}}{m^{2}c^{4}}}\right\} \tag{10.1}\label{10.1} 
\end{align}

%\enlargethispage{10\baselineskip}
where

\begin{align}
	\lambda & = \exp\Bigg[\frac{2mc^{2}\left(a-wt\right)}{\hbar\sqrt{c^{2}-w^{2}}}\Bigg] - \exp\Bigg[\frac{2mc^{2}(b-wt)}{\hbar\sqrt{c^{2}-w^{2}}}\Bigg], \tag{10.2.1}\label{10.2.1} \\
	\bar{\lambda} & = 2mc^{2} + A_{2}^{2}\hbar\sqrt{c^{2}-w^{2}} \Bigg\{\exp\Bigg[-\frac{2mc^{2}(b-wt)}{\hbar\sqrt{c^{2}-w^{2}}}\Bigg] \nonumber \\
	& - \exp\Bigg[-\frac{2mc^{2}\left(a-wt\right)}{\hbar\sqrt{c^{2}-w^{2}}}\Bigg]\Bigg\}. \tag{10.2.2}\label{10.2.2}
\end{align}
In \autoref{fig:10} and \autoref{fig:11}, we provide a graphical representation of the solution in \myeqref{9}. In \autoref{fig:10}, we observe a solitary wave with wave velocity $w<c$, and in \autoref{fig:11}, we show the density plot with $w>c$.

In the graphical representations, two cases can be identified: the example of a wave function that obeys causality and whose wave speed is less than $c$, and $\textit{vice versa}$. In the former case, an inverted cuspon that has a minimum value, $\Psi_{\textrm{min}} =0$, was attained for the particular choice of parameters (\autoref{fig:10}). In the latter case, the soliton solution that was purely imaginary, that is to say, it lay in the complex plane, was obtained. Here, the modulus of the wave function was graphed (\autoref{fig:11}). The chroma patterns alternate between dark and light, with increasing values being demarcated as dark.

%==============================================================================================

\section{The Nonlinear Klein-Gordon Equation}\label{TNLKGE}
\subsection{Soliton Solutions}\label{SS}
Traditionally, the nonlinear KG equation is obtained by including an additional spatial and temporally varying source term to the KG equation. Equations of this nature have piqued the interest of many physicists in the areas of quantum mechanics, quantum field theory \cite{ref16}, condensed matter and particle physics \cite{ref17}. We introduce the source term, $\alpha\Psi^{k}(\mathbf{r},t)$, where $\alpha, k \in \mathbb{R}$ are constants. \myeqref{6.2} becomes
\begin{equation}
	\frac{1}{c^{2}} \Psi_{tt}(\mathbf{r},t) - \nabla^{2}\Psi(\mathbf{r},t) + \dfrac{m^{2}c^{2}}{\hbar^{2}}\Psi(\mathbf{r},t) + \alpha\Psi^{k}(\mathbf{r},t) = 0. \tag{11.1}\label{11.1}
\end{equation}
In (1+1)-dimensions, \myeqref{11.1} reduces to
\begin{equation}
	\frac{1}{c^{2}} \Psi_{tt}(x,t) - \Psi_{xx}(x,t)\rangle + \dfrac{m^{2}c^{2}}{\hbar^{2}}\Psi(x,t) + \alpha\Psi^{k}(x,t) = 0. \tag{11.2}\label{11.2}
\end{equation} 
We pursue exact solutions to \myeqref{11.2} and provide general conservation laws to \myeqref{11.2}. Applying the transformations given in \myeqref{7.1} and \myeqref{7.2}, we obtain the reduced nonlinear ODE 
\begin{equation}
	\Big[\Big(\frac{w}{c}\Big) - 1 \Big] \dfrac{\mathrm{d}^{2}\phi}{\mathrm{d}{\Xi^{2}}} + \dfrac{m^{2}c^{2}}{\hbar^{2}}\phi + \alpha \phi^{k} = 0. \tag{12}\label{12}
\end{equation}
Using the sine-cosine \textit{ansatz} \cite{Wazwaz1}, we seek solutions of the form
\begin{equation}
	u(\Xi) = \lambda\cos^{\beta}(\chi\Xi), \quad \textrm{for} \quad \vert{\Xi}\vert \leq \dfrac{\pi}{2\chi}. \tag{13}\label{13}
\end{equation}
Substituting \myeqref{13}, and its associated derivatives, into \myeqref{12} and collecting terms by powers of $\cos(\chi\Xi)$, we obtain
\begin{align}
	&(c^{2}\hbar^{2}\beta\lambda\chi^{2} - \hbar^{2}w^{2}\beta\lambda\chi^{2} - c^{2}\hbar^{2}\beta^{2}\lambda\chi^{2} + \hbar^{2}w^{2}\beta^{2}\lambda\chi^{2}) \cos^{\beta}(\chi\Xi) \nonumber \\
	& + \Big(c^{2}\hbar^{2}\alpha\lambda^{k}\Big)\cos^{2+\beta k}(\chi\Xi) \hspace*{0.1cm} \times \nonumber \\
	& (c^{4}m^{2}\lambda + c^{2}\hbar^{2}\beta^{2}\lambda\chi^{2} - \hbar^{2}w^{2}\beta^{2}\lambda\chi^{2})\cos^{\beta + 2}(\chi\Xi) = 0. \tag{14.1}\label{14.1}
\end{align}
Balancing the powers and coefficients of $\cos(\chi\Xi)$, we obtain the following system of algebraic equations
\begin{align}
	& \beta = 2 + \beta k, \tag{15.2.1}\label{15.2.1} \\
	& c^{2}\hbar^{2}\beta\lambda\chi^{2} - \hbar^{2}w^{2}\beta\lambda\mu^{2} - c^{2}\hbar^{2}\beta^{2}\lambda\mu^{2} + \hbar^{2}w^{2}\beta^{2}\lambda\chi^{2} \nonumber \\
	& = c^{2}\hbar^{2}\alpha\lambda^{k}, \tag{15.2.2}\label{15.2.2} \\
	& c^{4}m^{2}\lambda + c^{2}\hbar^{2}\beta^{2}\lambda\chi^{2} - \hbar^{2}w^{2}\beta^{2}\lambda\chi^{2} = 0. \tag{15.2.3}\label{15.2.3}
\end{align}
Solving this system of equations, we obtain the triadic-solution
\begin{equation}
\hspace{-0.4cm}(\lambda,\beta,\chi) = \Bigg\{\Bigg(-\Bigg[\dfrac{2\hbar^{2}\alpha}{m^{2}c^{2}(k+1)}\Bigg]^{1/1-k}, -\dfrac{2}{k-1},\pm 
\dfrac{mc^{2}(k-1)}{2\hbar\sqrt{w^{2}-c^{2}}}\Bigg)\Bigg\}, \tag{16.1}\label{16.1}
\end{equation}
which leads to an expression for the field
\begin{equation}
\hspace{-0.5cm}	\Psi(x,t) = -^{k-1}\sqrt{\frac{2\hbar^{2}\alpha}{m^{2}c^{2}(k+1)}\sec^{2} \Bigg[\frac{mc^{2}(k-1)}{2\hbar\sqrt{w^{2}-c^{2}}}(x-wt)\Bigg]}, \tag{16.2}\label{16.2}
\end{equation}
with the restriction that $k \in \mathbb{N}-\left\{1\right\}$. We graphically demonstrate the solution in \myeqref{16.2} for the specific parameter values in \autoref{fig:12} below. The soliton in \autoref{fig:12} is of the cuspon-type, noting diverging values of the derivatives at the trough of the wave profile.

\begin{figure}[H]
	\centering
	\includegraphics[width=0.7\linewidth]{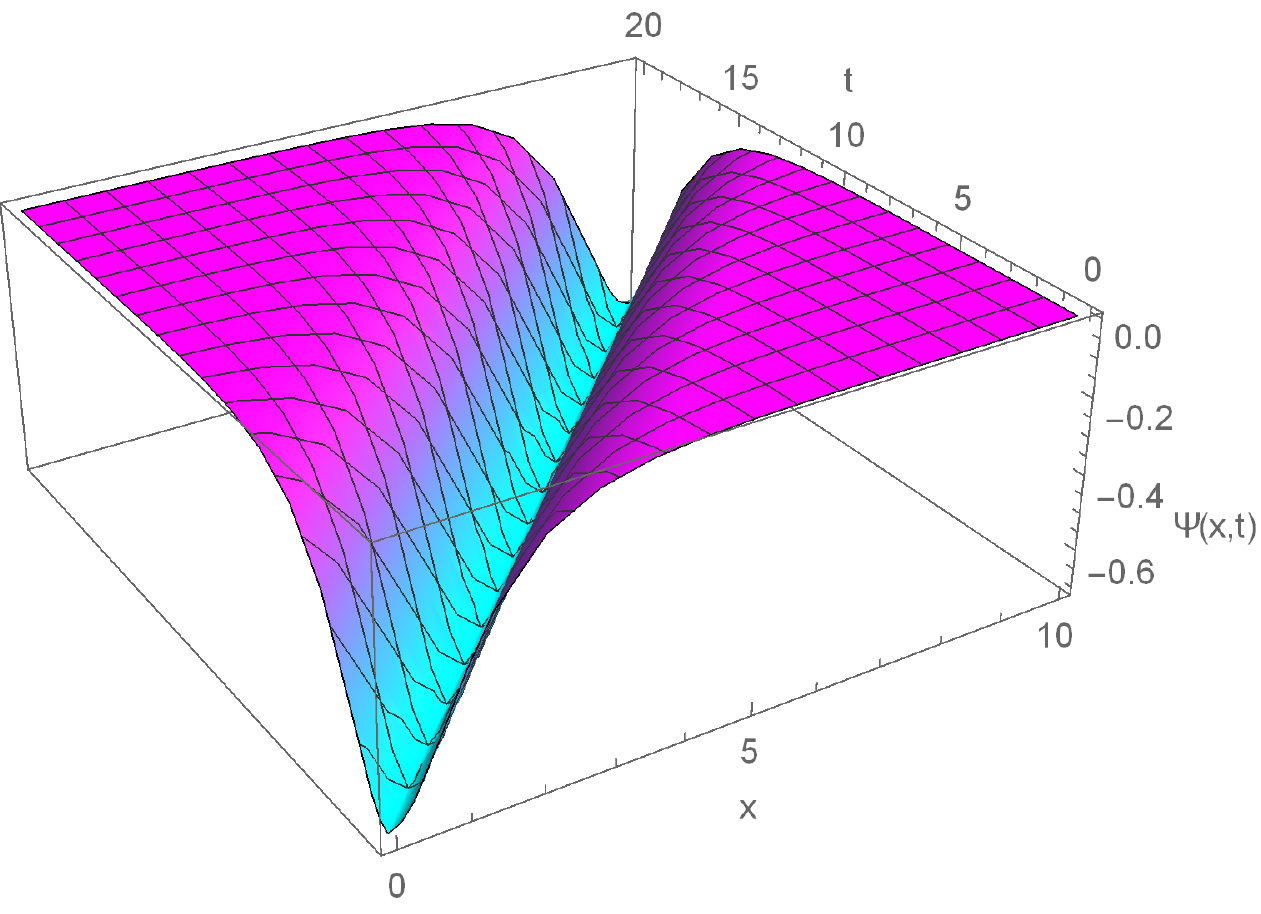}
	\caption{Plot of the one-soliton with $\alpha, m=1,$ and $k=3$ in relativistic units that obey causality with wave speed, $w=\frac{1}{2}$}
	\label{fig:12}
\end{figure}
For a particular parameter choice, an inverted-cuspon solution was obtained with a minimum value of $\Psi_{\textrm{min}} =0$.

\subsection{Conservation Laws}\label{CL}
Due to the linear KG equation being homogeneous and linear in the scalar field, the equation possesses a perpetual class of monotonous conservation laws, and any attempt at obtaining them would be futile. The conservation laws for the nonlinear KG equation proves to be more interesting. For the nonlinear KG equation, we consider multipliers of the form 
$\Lambda=\Lambda(x,y,z,t,\Psi_{x},\Psi_{y},\Psi_{z},\Psi_{t})$. The conservation laws for \myeqref{11.2} take the form
\begin{align}
	& \Lambda(x,y,z,t,\Psi_{x},\Psi_{y},\Psi_{z},\Psi_{t})\Bigg[\frac{1}{c^{2}}\Psi_{tt} - \nabla^{2}\Psi + \frac{m^{2}c^{2}}{\hbar^{2}}\Psi+\alpha\Psi^{k}\bigg] \nonumber \\
	& = D_{t}\Phi^{t} + D_{x}\Phi^{x} + D_{y}\Phi^{y} + D_{z}\Phi^{t}. \tag{17.1}\label{17.1}
\end{align}
Using $\texttt{MATHEMATICA}$ \cite{ref15}, we determine the general form of the multiplier to be 
\begin{align}
	\Lambda(x,y,z,t,\Psi_{x},\Psi_{y},\Psi_{z},\Psi_{t}) & = 	(\varsigma_{3}x+\varsigma_{5}y+\varsigma_{6}z+\varsigma_{10})\Psi_{t}  \nonumber \\ &+(\varsigma_{3}c^{2}t-\varsigma_{1}y-\varsigma_{2}z+\varsigma_{7})\Psi_{x} \nonumber \\
	&+\; (\varsigma_{5}c^{2}t+\varsigma_{1}x-\varsigma_{4}z+\varsigma_{8})\Psi_{y} \nonumber \\
	&+\; (\varsigma_{6}c^{2}t+\varsigma_{2}x+\varsigma_{4}y+\varsigma_{9})\Psi_{z}. \tag{17.2}\label{17.2}
\end{align}
where $\varsigma_{i} \in \mathbb{R}$ are constants, for $1\leq i \leq 10$. We summarize the conservation laws in \autoref{table:2} - \autoref{table:4} below. \\

We note that the conservation laws in \autoref{table:2} - \autoref{table:4} are so complex that it is not obvious what physical meaning we can attach to them. We resort to calling them \textit{exotic} conservation laws, which may be non-observable, but exist much like the case in orbital mechanics with the Laplace-Runge-Lenz vector. Additionally, if we set $\alpha=0$, we recover the known conservation laws for the KG equation in the literature.  

 \FloatBarrier
\begin{table}[p]
	\centering
	\caption{Multipliers and conservation laws for the (3+1)-dimensional KG equation with arbitrary $\alpha,k\in\mathbb{R}$ such that $k>1$  -- Part I}
	\vspace{0.1cm}
	\label{table:2}
	\renewcommand{\arraystretch}{2}
	\begin{tabular}{ll}
		\hline
		\hline
		\hspace*{0.5cm}\textbf{Multiplier} & \hspace*{4.5cm}\textbf{Conserved Currents} \\
		\hline
		$\Lambda_{1} = x\Psi_{y} - y\Psi_{x}$ & $\Phi_{1}^{t} = \alpha tx\Psi^{k}\Psi_{y} - \alpha ty\Psi^{k}\Psi_{x} - \frac{m^{2}c^{2}}{\hbar^{2}}tx\Psi\Psi_{y} + \frac{m^{2}c^{2}}{\hbar^{2}}ty\Psi\Psi_{x}  
		+ \frac{1}{c^{2}}y\Psi_{t}\Psi_{x} - \frac{1}{c^{2}}x\Psi_{t}\Psi_{y}$ \\ 
		& $\Phi_{1}^{x} = -\dfrac{m^{2}c^{2}}{\hbar^{2}}ty\Psi\Psi_{t} + \alpha ty\Psi^{k}\Psi_{t} + x\Psi_{x}\Psi_{y} - \frac{1}{2c^{2}}y\Psi^{2}_{t} - \frac{1}{2}y\Psi_{x}^{2} + \frac{1}{2}y\Psi_{y}^{2}+\frac{1}{2}y\Psi_{z}^{2}$ \\
		& $\Phi_{1}^{y} = -y\Psi_{x}\Psi_{y} + \dfrac{1}{2c^{2}}x\Psi_{t}^{2} + \frac{1}{2}x\Psi_{y}^{2} - \frac{1}{2}x\Psi_{x}^{2} - \frac{1}{2}x\Psi_{z}^{2} + \frac{m^{2}c^{2}}{\hbar^{2}}tc\Psi\Psi_{t} - \alpha tx\Psi^{k}\Psi_{t}$ \\
		& $\Phi_{1}^{z} = x\Psi_{y}\Psi_{z} - y\Psi_{x}\Psi_{z}$ \\
		\hline
		$\Lambda_{2} = x\Psi_{z} - z\Psi_{x}$ & $\Phi_{2}^{t} = -\alpha tz \Psi^{k}\Psi_{x} + \frac{m^{2}c^{2}}{\hbar^{2}}tz\Psi\Psi_{x} + \alpha tx \Psi^{k}\Psi_{z} - \frac{m^{2}c^{2}}{\hbar^{2}} tx\Psi\Psi_{z} +\dfrac{1}{c^{2}}z\Psi_{t}\Psi_{x} - \dfrac{1}{c^{2}}x\Psi_{t}\Psi_{z}$ \\ 
		& $\Phi_{2}^{x} = -\frac{m^{2}c^{2}}{\hbar^{2}}tz \Psi\Psi_{t} + \alpha tz\Psi^{k}\Psi_{t} + x\Psi_{x}\Psi_{z} -\dfrac{1}{2c^{2}}z\Psi_{t}^{2} -\frac{1}{2}z\Psi_{x}^{2} +\frac{1}{2}z\Psi_{y}^{2}+\frac{1}{2}z\Psi_{z}^{2}$\\
		& $\Phi_{2}^{y} = x\Psi_{y}\Psi_{z} - z\Psi_{x}\Psi_{y}$ \\
		& $\Phi_{2}^{z} = \dfrac{1}{2c^{2}}x\Psi_{t}^{2} - \frac{1}{2}x\Psi_{x}^{2} - \frac{1}{2}x\Psi_{y}^{2} + \frac{1}{2}x\Psi_{z}^{2} + \frac{m^{2}c^{2}}{\hbar^{2}}tx\Psi\Psi_{t} - \alpha tx\Psi^{k}\Psi_{t} - z\Psi_{x}\Psi{z}$\\
		\hline
		$\Lambda_{3} = c^{2}t\Psi_{x} + x\Psi_{t}$ & $\Phi_{3}^{t} = -\dfrac{1}{2c^{2}}x\Psi_{t}^{2} + \dfrac{1}{\hbar^{2}}\Big(\frac{\alpha\hbar^{2}}{k+1}x\Psi^{k+1} - \frac{m^{2}c^{2}}{2}x\Psi^{2}\Big) - \frac{1}{2}x\Psi_{x}^{2} - \frac{1}{2}x\Psi_{z}^{2} - \frac{m^{2}c^{2}}{2\hbar^{2}}t^{2}\Psi\Psi_{x}$ \\
		&\hspace*{1cm}$- t\Psi_{t}\Psi_{x} + \dfrac{\alpha c^{2}}{2}t^{2}\Psi^{k}\Psi_{x} - \frac{1}{2}x\Psi_{y}^{2}$ \\
		& $\Phi_{3}^{x} = \frac{1}{2}t\Psi_{t}^{2} - \dfrac{1}{2\hbar^{2}}(\alpha c^{2}\hbar^{2}t^{2}\Psi^{k} - m^{2}c^{4}t^{2}\Psi)\Psi_{t} + x\Psi_{t}\Psi_{x} + \frac{c^{2}}{2}t\Psi_{x}^{2} - \frac{c^{2}}{2}t\Psi_{y} -\frac{c^{2}}{2}t\Psi_{z}$ \\
		& $\Phi_{3}^{y} = c^{2}t\Psi_{x}\Psi_{y} + c\Psi_{t}\Psi_{y}$ \\
		& $\Phi_{3}^{z} = c^{2}t\Psi_{x}\Psi_{z} + x\Psi_{t}\Psi_{z}$ \\
		\hline
	\end{tabular}
\end{table}
\FloatBarrier

\FloatBarrier
\begin{table}[p]
	\centering
	\caption{Multipliers and conservation laws for the (3+1)-dimensional KG equation with arbitrary $\alpha,k\in\mathbb{R}$ such that $k>1$  -- Part II}
	\vspace{0.1cm}
	\label{table:3}
	\renewcommand{\arraystretch}{2}
	\begin{tabular}{ll}
		\hline
		\hline
		\hspace*{0.5cm}\textbf{Multiplier} & \hspace*{4.5cm}\textbf{Conserved Currents} \\
		\hline
		$\Lambda_{4} = y\Psi_{z} - z\Psi_{y}$ & $\Phi_{4}^{t} = -\alpha tz\Psi^{k}\Psi_{y} + \frac{m^{2}c^{2}}{\hbar^{2}} tz\Psi\Psi_{y} +\alpha ty\Psi^{k}\Psi_{z} - \frac{m^{2}c^{2}}{\hbar^{2}}ty\Psi\Psi_{z} + \frac{1}{c^{2}}z\Psi_{t}\Psi_{y} - \frac{1}{c^{2}}y\Psi_{t}\Psi_{z}$ \\
		& $\Phi_{4}^{x} = y\Psi_{x}\Psi_{z} - z\Psi_{x}\Psi_{y}$ \\
		& $\Phi_{4}^{y} = y\Psi_{y}\Psi_{z} - \dfrac{1}{2c^{2}}z\Psi_{t}^{2} - \frac{1}{2}z\Psi_{y}^{2} + \frac{1}{2}z\Psi_{x}^{2} + \frac{1}{2}z\Psi_{z}^{2} -\frac{m^{2}c^{2}}{\hbar^{2}}tz\Psi\Psi_{t} + \alpha tz \Psi^{k}\Psi_{t}$ \\
		& $\Phi_{4}^{z} = \dfrac{1}{2c^{2}}y\Psi_{t}^{2} -\frac{1}{2}y\Psi_{x}^{2} - \frac{1}{2}y\Psi_{y}^{2} + \frac{1}{2}y\Psi_{z}^{2} +\frac{m^{2}c^{2}}{\hbar^{2}}ty\Psi\Psi_{t} - \alpha ty\Psi^{k}\Psi_{t} - z\Psi_{y}\Psi_{z}$ \\
		\hline
		$\Lambda_{5} = c^{2}t\Psi_{y} + y\Psi_{t}$	& $ \Phi_{5}^{t} = -\dfrac{1}{2c^{2}}y\Psi_{t}^{2} + \frac{1}{\hbar^{2}}\Big(\frac{\alpha\hbar^{2}\Psi^{k+1}}{k+1} - \dfrac{m^{2}c^{2}}{2}y\Psi^{2}\Big) - \frac{1}{2}y\Psi_{x}^{2} - \frac{1}{2}y\Psi_{z}^{2} - \frac{m^{2}c^{4}}{2\hbar^{2}}t^{2}\Psi\Psi_{y}$ \\ &$\hspace*{1cm} - t\Psi_{t}\Psi_{y} + \dfrac{\alpha c^{2}}{2}t^{2}\Psi^{k}\Psi_{y} - \frac{1}{2}y\Psi_{y}^{2}$ \\
		& $\Phi_{5}^{x} = c^{2}t\Psi_{x}\Psi_{y} + y\Psi_{t}\Psi_{x}$ \\
		& $\Phi_{5}^{y} = y\Psi_{t}\Psi_{y} + \frac{c^{2}}{2}t\Psi_{y}^{2} -\frac{c^{2}}{2}t\Psi_{x}^{2} - \frac{c^{2}}{2}t\Psi_{z}^{2} + \frac{1}{2}t\Psi_{t}^{2} + \dfrac{1}{2\hbar^{2}}\Big(m^{2}c^{4}t^{2}\Psi - \alpha c^{2}\hbar^{2}t^{2}\Psi^{k}\Big)\Psi_{t}$ \\
		& $\Phi_{5}^{z} = c^{2}t\Psi_{y}\Psi_{z} + y\Psi_{t}\Psi_{z}$ \\
		\hline
		$\Lambda_{6} = c^{2}t\Psi_{z} + z\Psi_{t}$ & $\Phi_{6}^{t}  = -\dfrac{1}{2c^{2}}z\Psi_{t}^{2} + \frac{1}{\hbar^{2}} \Big(\frac{\alpha\hbar^{2}z\Psi^{k+1}}{k+1} - \dfrac{m^{2}c^{2}}{2}z\Psi^{2}\Big) - \frac{1}{2}z\Psi_{x} - \frac{1}{2}z\Psi^{2}z -t\Psi_{t}\Psi_{z}$ \\ &$\hspace*{1cm}- \frac{m^{2}c^{4}}{2\hbar^{2}}t^{2}\Psi\Psi_{z} + \dfrac{\alpha c^{2}}{2}t^{2}\Psi^{k}\Psi_{z} - \frac{1}{2}z\Psi_{y}^{2}$ \\
		& $\Phi_{6}^{x} = c^{2}t\Psi_{x}\Psi_{z} + z\Psi_{t}\Psi_{x}$ \\
		& $\Phi_{6}^{y} = c^{2}t\Psi_{y}\Psi_{z} + z\Psi_{t}\Psi_{y}$ \\
		& $\Phi_{6}^{z} = \frac{c^{2}}{2}t\Psi_{z}^{2} - \frac{c^{2}}{2}t\Psi_{x}^{2} - \frac{c^{2}}{2}t\Psi_{y}^{2} + \frac{1}{2}t\Psi_{t}^{2} + z\Psi_{t}\Psi_{z} - \dfrac{1}{2\hbar^{2}}\Big(\alpha c^{2}\hbar^{2}t^{2}\Psi^{k} - m^{2}c^{4}t^{2}\Psi\Big)$ \\
		\hline
	\end{tabular}
 \end{table}
\FloatBarrier

\FloatBarrier
\begin{table}[p]
	\centering
	\caption{Multipliers and conservation laws for the (3+1)-dimensional KG equation with arbitrary $\alpha,k\in\mathbb{R}$ such that $k>1$ -- Part III}
	\vspace{0.1cm}
	\label{table:4}
	\renewcommand{\arraystretch}{1.7}
	\begin{tabular}{ll}
		\hline
		\hline
		\textbf{Multiplier} & \hspace*{4.5cm}\textbf{Conserved Currents} \\
		\hline
		$\Lambda_{7} = \Psi_{x}$ & $\Phi_{7}^{t} = -\frac{1}{c^{2}}\Psi_{t}\Psi_{x} + \alpha t\Psi^{k}\Psi_{x} -\dfrac{m^{2}c^{2}}{\hbar^{2}}t\Psi\Psi_{x}$ \\ 
		& $\Phi_{7}^{x} = \frac{1}{2}\Psi_{x}^{2} - \frac{1}{2}\Psi_{y}^{2} - \frac{1}{2}\Psi_{z}^{2} + \frac{1}{2c^{2}}\Psi_{t}^{2} -\alpha t\Psi^{k}\Psi_{t} + \frac{m^{2}c^{2}t\Psi\Psi_{t}}{\hbar^{2}}$ \\
		& $\Phi_{7}^{y} = \Psi_{x}\Psi_{y}$ \\
		& $\Phi_{7}^{z} = \Psi_{x}\Psi_{z}$ \\
		\hline
		$\Lambda_{8} = \Psi_{y}$ & $\Phi_{8}^{t} = -\frac{1}{c^{2}}\Psi_{t}\Psi_{y} - \frac{m^{2}c^{2}t\Psi\Psi_{y}}{\hbar^{2}} + \alpha t\Psi^{k}\Psi_{y}$ \\ 
		& $\Phi_{8}^{x} = \Psi_{x}\Psi_{y}$ \\
		& $\Phi_{8}^{y} = \frac{1}{2}\Psi_{y}^{2} - \frac{1}{2}\Psi_{x}^{2} - \frac{1}{2}\Psi_{z}^{2} + \dfrac{1}{2c^{2}}\Psi_{t}^{2} + \frac{m^{2}c^{2}t\Psi\Psi_{t}}{\hbar^{2}} - \alpha t \Psi^{k}\Psi_{t}$ \\
		& $\Phi_{8}^{z} = \Psi_{y}\Psi_{z}$ \\
		\hline
		$\Lambda_{9} = \Psi_{z}$ & $\Phi_{9}^{t} = -\frac{1}{c^{2}}\Psi_{t}\Psi_{z} - \frac{m^{2}c^{2}t\Psi\Psi_{z}}{\hbar^{2}} + \alpha t\Psi^{k}\Psi_{z}$ \\
		& $\Phi_{9}^{x} = \Psi_{x}\Psi_{z}$ \\
		& $\Phi_{9}^{y} = \Psi_{y}\Psi_{z}$ \\
		& $\Phi_{9}^{z} = -\frac{1}{2}\Psi_{x}^{2} - \frac{1}{2}\Psi_{y}^{2} + \frac{1}{2}\Psi_{z}^{2} + \dfrac{1}{2c^{2}}\Psi_{t}^{2} - \alpha t \Psi^{k}\Psi_{t} + \frac{m^{2}c^{2}t\Psi\Psi_{t}}{\hbar^{2}}$ \\
		\hline
		$\Lambda_{10} = \Psi_{t}$ &  $\Phi_{10}^{t} = -\frac{1}{2}\Psi_{x}^{2} - \frac{1}{2}\Psi_{z}^{2} + \dfrac{1}{\hbar^{2}}\Big(\frac{\alpha\hbar^{2}\Psi^{k+1}}{k+1} - \frac{m^{2}c^{2}}{2}\Psi^{2}\Big) - \frac{1}{2}\Psi_{y}^{2} - \dfrac{1}{2c^{2}}\Psi_{t}^{2}$ \\
		& $\Phi_{10}^{x} = \Psi_{t}\Psi_{x}$ \\
		& $\Phi_{10}^{y} = \Psi_{t}\Psi_{y}$ \\
		& $\Phi_{10}^{z} = \Psi_{t}\Psi_{z}$ \\
		\hline
	\end{tabular}
\end{table}
\FloatBarrier

%===============================================================================================
\clearpage
\section{Conclusion}\label{CONCLUSION}
In this paper, solution methodologies for the linear KG equation were explored in terms of Lie point symmetries. The solutions obtained, via similarity reductions, proved impotent in this regard. However, a linear combination of spatial and temporal point symmetries demonstrated to be beneficial, and a soliton solution, which was normalizable, was obtained. \\ 

Thereafter, a study of the nonlinear KG equation, which arises due to the addition of a source term to the linear KG equation, was conducted. This source term need only depend on spacetime arbitrarily, and in this study, it was chosen to be proportional to an arbitrary power of the scalar field itself. \\

Soliton solutions to the equation were obtained using the sine-cosine method. A new solution was obtained in terms of a generalized radical function, which was normalizable. \\

Lastly, conservation laws, applied to the $(3+1)$-dimensional nonlinear KG equation, were obtained using the multiplier method. Ten new, non-trivial conservation laws emanated. It is interesting to note that if the coefficient, $\alpha$, vanishes, conservation laws of the linear KG equation are obtained. Furthermore, we have the freedom of electing any integer value greater than unity for the parameter $\textit{k}$ and thus, the conservation laws can be obtained algorithmically. 

%================================================================================================

\section{Appendix}\label{APPENDIX}
\subsection{Lie Point Symmetries}
Following Olver \cite{ref18}, consider the $m^{\text{th}}$-order PDE
\begin{equation}
	\Pi(t,x,\Psi,\Psi_{t},\Psi_{x},\Psi_{tt},\Psi_{xx},\ldots)=0, \tag{A.1}\label{A.1}
\end{equation} 
where $\Psi=\Psi(t,x)$. \myeqref{A.1} admits Lie point symmetries of the form
\begin{equation}
	\Omega=\tau(t,x,\Psi)\partial_{t}+\xi(t,x,\Psi)\partial_{x}+\eta(t,x,\Psi)\partial_{\Psi}. \tag{A.2}\label{A.2}
\end{equation}
The $2^{\text{nd}}$-order prolongation, which are natural extensions of the one-parameter Lie group of transformations, on the jet space $\left(t,x,\Psi,\Psi_{t},\Psi_{x}\right)$, is given by
\begin{align}
	\text{pr}^{\left[2\right]}\Omega=\Omega^{\left[2\right]}=&\;\Omega+\zeta^{\left[t\right]}\partial_{\Psi_{t}}+\zeta^{\left[x\right]}\partial_{\Psi_{x}}+\zeta^{\left[tt\right]}\partial_{\Psi_{tt}}+\zeta^{\left[xx\right]}\partial_{\Psi_{xx}}, \tag{A.3.1}\label{A.3.1} \\
	\zeta^{\left[t\right]}=&\;D_{t}\left(\eta\right)-D_{t}\left(\tau\right)\Psi_{t}-D_{t}\left(\xi\right)\Psi_{x}, \tag{A.3.2}\label{A.3.2} \\
	\zeta^{\left[x\right]}=&\;D_{x}\left(\eta\right)-D_{x}\left(\tau\right)\Psi_{t}-D_{x}\left(\xi\right)\Psi_{x}, \tag{A.3.3}\label{A.3.3} \\
	\zeta^{\left[tt\right]}=&\;D_{t}\left(\zeta^{\left[t\right]}\right)-D_{t}\left(\tau\right)\Psi_{tt}-D_{t}\left(\xi\right)\Psi_{xt}, \tag{A.3.4}\label{A.3.4} \\
	\zeta^{\left[xx\right]}=&\;D_{x}\left(\zeta^{\left[x\right]}\right)-D_{x}\left(\tau\right)\Psi_{tx}-D_{x}\left(\xi\right)\Psi_{xx}, \tag{A.3.5}\label{A.3.5} 
\end{align}
where $D_{i}$, for $i\in\left\{t,x\right\}$, is the total differential operator defined as
\begin{align}
&	D_{i}:=\frac{\partial}{\partial x^{i}}+\Psi_{i}^{\alpha}\frac{\partial}{\partial \Psi^{\alpha}}+\Psi^{\alpha}_{ij}\frac{\partial}{\partial \Psi^{\alpha}_{j}}+\ldots+\Psi^{\alpha}_{ii_{1}i_{2}\ldots i_{\ell}}\frac{\partial}{\partial \Psi_{i_{1}i_{2}\ldots i_{\ell}}^{\alpha}}+\ldots, \tag{A.4}\label{A.4} \\
& i=1,2,\ldots, \ell. \nonumber
\end{align}
The application of \myeqref{A.3.1} on \myeqref{A.1} yield a system of linear PDEs
\begin{equation}
	\tau\frac{\partial\Pi}{\partial t}+\xi\frac{\partial\Pi}{\partial x}+\eta\frac{\partial\Pi}{\partial\Psi}+\zeta^{\left[t\right]}\frac{\partial\Pi}{\partial\Psi_{t}}+\zeta^{\left[x\right]}\frac{\partial\Pi}{\partial\Psi_{x}}+\zeta^{\left[tt\right]}\frac{\partial\Pi}{\partial\Psi_{tt}}+\zeta^{\left[xx\right]}\frac{\partial\Pi}{\partial\Psi_{xx}}=0. \tag{A.5}\label{A.5}
\end{equation}
Equating the various powers of the variable, $\Psi$, in \myeqref{A.5} to zero produces a set of linear PDEs. Solving the system of PDEs provide functional forms for $\tau,\xi$, and $\eta$, which generate each Lie point symmetry.

\subsection{The Multiplier / Direct Method}
For a system of $m$ independent variables, we define the Euler-Lagrange derivative
\begin{align}
	&\frac{\delta}{\delta u^{\alpha}}:=\frac{\partial}{\partial u^{\alpha}}+\sum_{r\geq 1}\left(-1\right)^{r}D_{i_{1}}, D_{i_{2}},\ldots,D_{i_{r}}\frac{\partial}{\partial u^{\alpha}_{i_{1}i_{2}\ldots i_{r}}}, \tag{B.1}\label{B.1} \\ 
	&\alpha=1,2,\dots,m. \nonumber
\end{align}
The derivative defined in \myeqref{B.1} is a generalization of the derivative operator in the Euler-Lagrange equations used in mechanics. The multiplier method was proposed by Anco and Bluman \cite{ref19, ref20, ref21} as a generalization of Noether's theorem. We state the method for obtaining the conservation laws by the following recipe: 
\begin{enumerate}
 \item For a given system of PDEs, \newline
 $F_{m}\left(\mathbf{x},u,\partial u, \partial^{2}u,\ldots,\partial^{k}u\right)=0$, where 
	\begin{equation}
\hspace*{-0.5cm}	\partial^{p}=\left\{\frac{\partial^{p}u^{\mu}(\mathbf{x})}{\partial x_{i_{1}}\partial x_{i_{2}}\ldots\partial_{i_{p}}}\;\bigg|\;\mu=1,2,\ldots, n; i_{1},i_{2}\ldots, i_{p}=1,2,\ldots, q\right\}, \tag{B.2.1}\label{B.2.1}
	\end{equation}
	we seek multipliers of the form
	\begin{equation}
		\left\{\Lambda_{m}\left(\mathbf{x},u,\partial u,\partial^{2}u,\ldots,\partial^{\ell}u\right)\right\}_{m=1}^{q}, \tag{B.2.2}\label{B.2.2}
	\end{equation}
	for some specified order, $\ell$. It is imperative that we choose the dependence of the multipliers in such a manner that no singular multipliers arise.
 \item Apply the Euler-Lagrange derivative in \myeqref{B.1} to the product of the multiplier and the system of PDEs in order to obtain a set of determining equations. Thereafter, solve the set of determining equations, so as to determine all the multipliers. 
 \item Find the corresponding fluxes that satisfy the identity
	\begin{equation}
\hspace{-0.4cm}	\Lambda_{m}\left(\mathbf{x},u,\partial u,\partial^{2}u,\ldots,\partial^{\ell}u\right)F_{m}\left(\mathbf{x},u,\partial u,\partial^{2}u,\ldots,\partial^{k}u\right)=D_{i}\varphi^{i}. \tag{B.3}\label{B.3}
	\end{equation}
 \item Each set of fluxes and multipliers yield a local conservation law of the form
	\begin{equation}
		D_{i}\varphi^{i}\left(\mathbf{x},u,\partial u,\partial^{2}u,\ldots, \partial^{r}u\right)=0. \tag{B.4}\label{B.4}
	\end{equation}
\end{enumerate} 

\subsection{The Sine-cosine Method}
The sine-cosine method was introduced by Wazwaz \cite{ref22, ref23} for finding travelling wave solutions of generalized Korteweg-de Vries-type (KdV-type) equations. Wadati \cite{Wadati1, Wadati2, Wadati3} defines a \textit{soliton} as a nonlinear wave that is localized and propagates without changing its physical properties (shape,
velocity, frequency, amplitude, wavelength, etc.). It also has a particle property; it is a localized wave that has infinite support or
exponential tails. Furthermore, solitons are waves that are stable against mutual collisions, and they retain their identity.

Below we give the recipe for finding soliton solutions using the sine-cosine method:
\begin{enumerate}
	\item Consider the system of PDEs of the form 
	\begin{equation}
		\mathcal{F}_{m}(u_{t},u_{x},u_{tt},u_{xx},u_{tx},\ldots)=0. \tag{C.1}\label{C.1}
	\end{equation}
	Introduce the variable
	\begin{equation}
		\Xi=\mathbf{x}-wt, \tag{C.2}\label{C.2}
	\end{equation}
	where $w<\infty$ is the velocity of the wave profile. Therefore, the system in \myeqref{C.1} becomes
	\begin{equation}
		\mathcal{G}_{m}\left(u,\frac{du}{d\Xi},\frac{d^{2}u}{d\Xi^{2}},\ldots\right)=0. \tag{C.3}\label{C.3}
	\end{equation}
	\item Consider solutions to \myeqref{C.3} of the form
	\begin{equation}
		u(\Xi)=
		\begin{cases}
			\lambda\cos^{\beta}\left(\chi\Xi\right), \quad\quad |\Xi\leq\frac{\pi}{2\chi}, \\
			\lambda\sin^{\beta}\left(\chi\Xi\right), \quad\quad |\Xi|\leq\frac{\pi}{\chi},
		\end{cases} \tag{C.4}\label{C.4}
	\end{equation}
	where $\lambda, \beta$, and $\chi$ are constants to be determined.
	\item Take the associated derivatives of \myeqref{C.4} and substitute them into the system of ODEs in \myeqref{C.3} while making use of the trigonometric identity $\sin^{2}\left(\chi\Xi\right)+\cos^{2}\left(\chi\Xi\right)=1.$
	\item Group the resulting expressions according to the powers of $\sin\left(\chi\Xi\right)$ or $\cos\left(\chi\Xi\right)$. Thereafter, balance the powers and equate the coefficients to obtain a system of nonlinear algebraic equations.
	\item Solve the resulting system of equations and reintroduce the original variables in order to attain the solution. 
\end{enumerate}

\section*{Acknowledgements}\label{ACKNOWLEDGEMENTS}
\noindent MAZK, MM, and FP thank P. G. L. Leach for a thorough review of the material, his helpful comments, and suggestions. MAZK and FP acknowledges that this research is supported by the South African Research Chair Initiative of the Department of Science and Innovation and the National Research Foundation (NRF).


\newpage 
\section{Bibliography}\label{BIBLIOGRAPHY}

\begin{thebibliography}{}
	\bibitem{ref1} Ohlsson T 2011 \textit{Relativistic Quantum Physics - From Advanced Quantum Mechanics to Introductory Quantum Field Theory} (Cambridge: Cambridge University Press).
	
	\bibitem{QFT1} Casalbuoni R 2011 \textit{Introduction to Quantum Field Theory} (Singapore: World Scientific Publishing Company). 
	
	\bibitem{QFT2} Sakurai J J, Napolitano J 2020 \textit{Modern Quantum Mechanics} (Cambridge: Cambridge University Press).
	
	\bibitem{ref3} Weinberg S 2002 \textit{The Quantum Theory of Fields} (Cambridge: Cambridge University Press).
	
	\bibitem{Klein} Klein O 1926 \textit{Z. Phys.} \textbf{37} 875.
	
	\bibitem{Gordon} Gordon W 1926 \textit{Z. Phys.} \textbf{40} 1-2.
	
	\bibitem{Schroedinger} Schr\"{o}dinger E 1926 \textit{Ann. Phys.} \textbf{81} 109.
	
	\bibitem{ref14} Lancaster T, Blundell S J 2014 \textit{Quantum Field Theory for the Gifted Amateur} (Oxford: Oxford University Press).
	
	\bibitem{ref2} Srednicki M 2007 \textit{Quantum Field Theory} (Cambridge: Cambridge University).
	
	\bibitem{ref5} Peskin M E, Schroeder D V 2019 \textit{An Introduction to Quantum Field Theory} (Florida: CRC Press).
	
	\bibitem{ref16} Bulbul B, Sezer M, Greiner W 2000 \textit{Relativistic Quantum Mechanics - Wave Equations} (Berlin: Springer).
	
	\bibitem{Paliathanasis} Paliathanasis A, Tsamparlis M, Mustafa M T 2015 \textit{Int. J. Geo. Met. Mod. Phys.} \textbf{12} 3.
	
	\bibitem{JamalKaraBokhari} Jamal S, Kara A H, Bokhari A H 2012 \textit{Can. J. Phys.} \textbf{90} 7.
	
	\bibitem{Azad} Azad H, Al-Dweik A Y, Ghanam R, Mustafa M T 2013 \textit{J. Math. Phys.} \textbf{54} 063509.
	
	\bibitem{PaliathanasisTsamparlis}  Paliathanasis A, Tsamparlis M 2014 \textit{Int. J. Geo. Met. Mod. Phys.} \textbf{11} 4.
	
	\bibitem{Inc} Inc M, Yusuf A, Aliyu A I, Baleanu D 2018 \textit{Phys. A: Stat. Mech. Appl.} \textbf{493} 94-106.
	
	\bibitem{Biswas} Biswas A, Kara A H, Bokhari A H, Zaman F D 2013 \textit{Non. Dyn.} \textbf{73} 4.
	
	\bibitem{KhalqiueBiswas} Khalique C M, Biswas A 2010 \textit{App. Math. Lett.} \textbf{23} 11.
	
	\bibitem{Wazwaz1} Wazwaz A-M 2005 \textit{App. Math. Comp.} \textbf{167} 2.		
	
	\bibitem{ref11} Gravel P, Gauthier C 2011 \textit{Am. J. Phys.} \textbf{79} 5.
	
	\bibitem{ref12} Greiner W 2000 \textit{Relativisitic Quantum Mechanics. Wave Equations}; (Berlin: Springer).
	
	\bibitem{Kragh} Kragh H 1984 \textit{Am. J. Phys.} \textbf{52} 11.
	
	\bibitem{Pauli} Pauli W 1979 \textit{Wissenschaftlicher Briefwechsel} eds A. Hermann, K. v. Meyenn, and V. F. Weisskodf (New York: Springer: New York). 
	
	\bibitem{ref15} Wolfram S 2007 \textit{The Mathematica Book} (Champaign: Wolfram Media Inc.). 
	
	\bibitem{ref22} Wazwaz A-M 2004 \textit{Math. Comput. Modell.} \textbf{40} 5-6.
	
	\bibitem{ref23} Wazwaz A-M 2004 \textit{Math. Comput. Modell.} \textbf{159} 2.	
	
	\bibitem{ref17} Caudrey P J, Eilbeck J C, Gibbon J D 1975 \textit{I. Nuo. Cim. B Ser. 11} \textbf{25} 2. 
	
	\bibitem{ref18} Olver P 1986 \textit{Applications of Lie Groups to Differential Equations} (Berlin: Springer).
	
	\bibitem{ref19} Anco S C, Bluman G W 1997 \textit{Phys. Rev. Lett.} \textbf{78} 5.
	
	\bibitem{ref20} Anco S C, Bluman G W 2002 \textit{Euro. J. Appl. Math.} \textbf{13} 05.
	
	\bibitem{ref21} Anco S C, Bluman G W 2002 \textit{Euro. J. Appl. Math.} \textbf{13} 5.
	
	\bibitem{Wadati1} Wadati M 2001 \textit{J. Phys.} \textbf{57} 5-6.
	
	\bibitem{Wadati2} Wadati M 1972 \textit{J. Phys. Jpn.} \textbf{32} 5.
	
	\bibitem{Wadati3} Wadati M 1973 \textit{J. Phys. Jpn.} \textbf{34} 5.
	
\end{thebibliography}
\end{document}